\documentclass[aps,prb,twocolumn,groupaddress,floatfix]{revtex4}

\usepackage{graphicx}
\usepackage{dcolumn}
\usepackage{bm}
\usepackage{amsmath}
\usepackage{multirow}

\newcommand{\tbh}[3]{\multicolumn{#1}{#2}{\textbf{#3}}}
\newcommand{\tbhc}[1]{\multicolumn{1}{c}{\textbf{#1}}}
\newcommand{\tbhl}[1]{\multicolumn{1}{l}{\textbf{#1}}}

\newcolumntype{f}[1]{D{.}{.}{#1}}

\begin{document}

\title{ Theoretical X-Ray Absorption Debye-Waller Factors}
\date{\today}

\author{Fernando D. Vila and J. J. Rehr}
\affiliation{Department of Physics, University of Washington, Seattle, WA 98195}
\author{H. H. Rossner and H. J. Krappe}
\affiliation{Hahn-Meitner-Institut Berlin, Glienicker Strasse 100, D-14109 Berlin, Germany}

\begin{abstract} 
An approach is presented for  theoretical
calculations of the Debye-Waller factors in x-ray absorption spectra.
These factors are represented in terms
of the cumulant expansion up to third order. They account respectively
for the net thermal expansion $\sigma^{(1)}(T)$,
the mean-square relative displacements $\sigma^2(T)$, and
the asymmetry of the pair distribution function $\sigma^{(3)}(T)$.
Similarly, we obtain Debye-Waller factors for x-ray and neutron scattering
in terms of the mean-square vibrational amplitudes $u^2(T)$.
Our method is based on density functional theory calculations
of the dynamical matrix,
together with an efficient Lanczos algorithm for projected
phonon spectra within the quasi-harmonic approximation.
Due to anharmonicity in the interatomic forces, the results
are highly sensitive to variations in the equilibrium
 lattice constants, and hence to the choice of exchange-correlation
potential. In order to treat this sensitivity, we introduce two prescriptions:
one based on the local density approximation, and a second based on a
modified generalized gradient approximation.
Illustrative results for the leading cumulants are presented for several
materials and compared with experiment and with correlated Einstein
and Debye models.
We also obtain Born-von Karman parameters and corrections due to
perpendicular vibrations.
\end{abstract}

\date{\today}

\maketitle

\section{Introduction}
\label{sec:intro}


Thermal vibrations and disorder in x-ray absorption spectra (XAS)
give rise to Debye-Waller (DW) factors varying as $\exp[-W(T)]$,
where $W(T) \approx 2 k^2 \sigma^2(T)$ and $\sigma^2(T)$ is the
mean square relative displacement (MSRD) of a given multiple-scattering (MS)
path.\cite{crozier88}
These Debye-Waller factors damp the spectra with respect to increasing
temperature $T$ and
wave number $k$ (or energy), and account for the
observation that the x-ray absorption fine structure (XAFS),
``melts" with increasing temperature.\cite{shmidt63}
The XAFS DW factor is analogous to that for x-ray and neutron
diffraction or the M\"o{\ss}bauer effect, where $W(T)=(1/2) k^2 u^2(T)$.
The difference is that the XAFS
DW factor refers to correlated averages over relative displacements,
e.g., $\sigma^2=\langle[({\mathbf u}_R -{\mathbf u}_0)\cdot
{\hat{\mathbf R}}]^2\rangle$ for the MSRD,
while that for x-ray and neutron diffraction refers to
the mean-square displacements
$u^2(T)=\langle ({\bf u\cdot\hat{\bf R}}{})^2\rangle$
of a given
atom.  Due to their exponential damping, accurate
DW factors are crucial to a quantitative treatment of x-ray absorption
spectra.
Consequently, the lack of precise Debye-Waller factors has been 
one of the biggest limitations to accurate structure determinations
(e.g., coordination number and interatomic distances) from XAFS experiment.

Due to the difficulty of calculating the vibrational distribution
function from first principles, XAFS Debye-Waller factors have, heretofore,
been fitted to experimental data or estimated
semi-empirically, e.g., from correlated Einstein and Debye models.
\cite{sevillano79,hung97}
However, these {\it ad hoc} approaches are unsatisfactory for several
reasons. First, there are often many more DW factors in the MS
path expansion than can be fit reliably.  Second, semi-empirical models
typically ignore anisotropic contributions and hence do not capture the
detailed structure of the phonon spectra.

To address these problems, we introduce first principles procedures
for calculations of the
Debye-Waller factors in XAS and related spectra.  Our approach is  based
primarily on
density functional theory (DFT) calculations of the dynamical matrix,
together with an efficient Lanczos
algorithm for the projected phonon spectra.\cite{poiarkova01,krappe02}
DFT calculations of 
crystallographic Debye-Waller factors and other thermodynamic quantities
have been carried out previously using modern electronic structure codes,
\cite{baroni01,lee95,gonze96}
and our work here builds on these developments,
with particular emphasis on applications to XAS.

Due to intrinsic anharmonicity in the interatomic forces,
the behavior of the DW factors is extremely sensitive to
the equilibrium lattice constant $a$. For example, we find that 
$\sigma^2$ varies approximately as 
$a^{6\gamma}$, where $\gamma = - d\ln \bar\omega/d\ln V$ is the mean
Gr\"uneisen parameter which is typically about 2 for fcc metals,
and $\bar\omega$ refers to the mean phonon frequency.
Consequently $\sigma^2$ is also very sensitive to the  choice of the
exchange-correlation potential in the DFT, since a $1\%$ error in
lattice constant yields an error of
$6\gamma \approx 10\%$ in $\sigma^2$.  As a result, relatively small
errors in the lattice constant predicted by
the local density approximation (LDA) which tends to overbind, or the
generalized gradient approximation (GGA) which tends to underbind,
become greatly magnified \cite{narasimhan02} in DW calculations.

In order to treat this sensitivity
we have developed two {\it ad hoc}
prescriptions for {\it ab initio} calculations of Debye-Waller factors
based on DFT calculations with I) the conventional LDA
and II) a modified-GGA (termed hGGA) described below.
 For comparison we also present selected results with a
conventional GGA, with the correlated Einstein and Debye models, and with
an empirical model based on the Born-von Karman parameters obtained from
fits to phonon spectra. Detailed results are
presented for a number of fcc and diamond structures.



%

\section{Formalism}
\label{sec:meth}

\subsection{Cumulants}
\label{subsec:cumulants}

\begin{figure}[t]
\includegraphics[scale=0.35,clip]{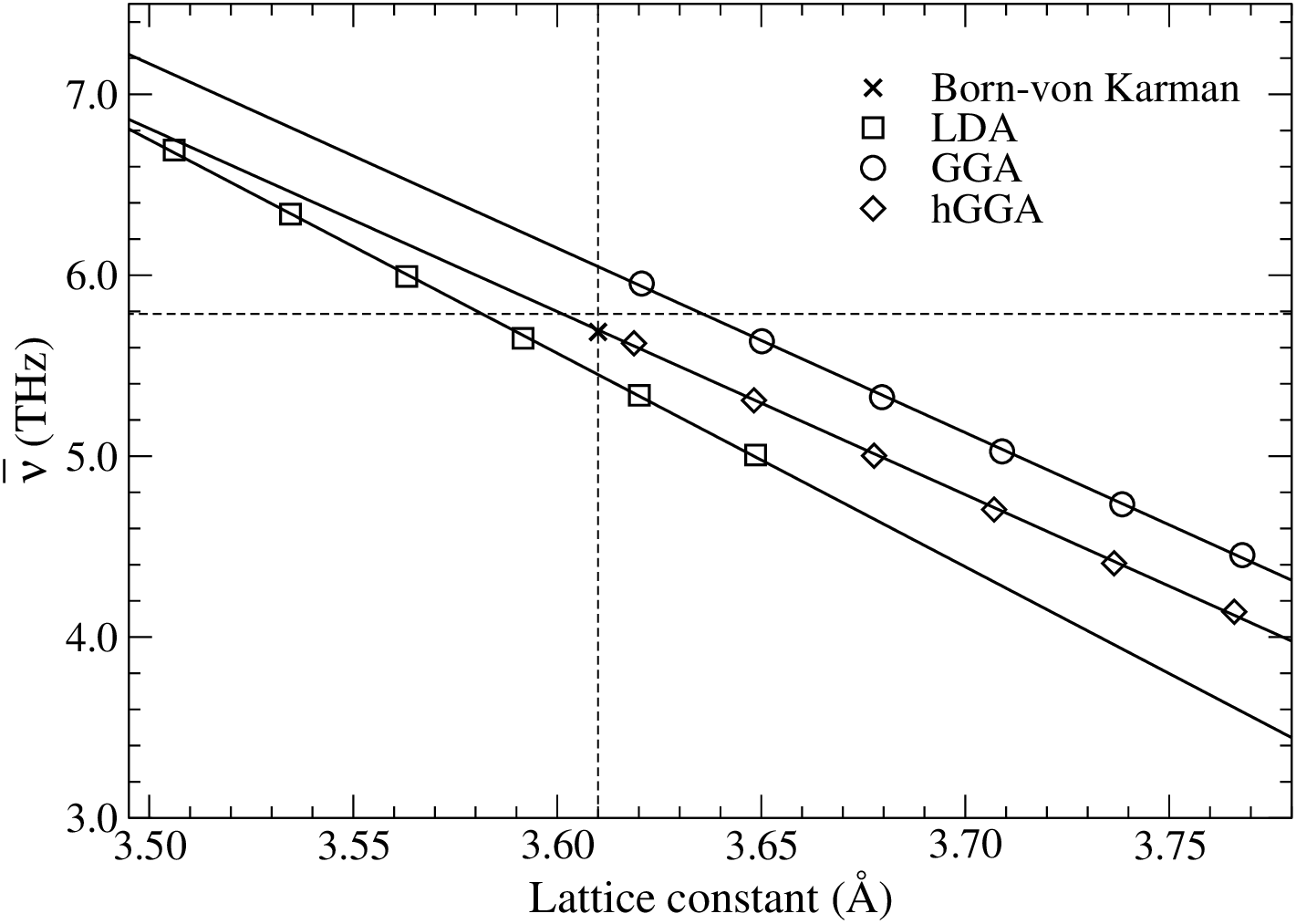}
\caption{\label{fig:cu_wavg}
Mean frequency $\bar\nu$ of the VDOS projected along the nearest neighbor single scattering 
path of Cu, obtained from the first Lanczos iteration.
The vertical line indicates the experimental lattice constant at 298 K while the
horizontal line shows the Einstein frequency obtained from the experimentally determined DW factor.
The Born-von Karman parameters for Cu at 298 K
were taken from Ref. \onlinecite{nicklow67}.
}
\end{figure}

In this section we outline the formalism used in our approach. 
Physically, the DW factors in XAS arise from 
a thermal and configurational average of
the XAS spectra $\langle \mu(E)\rangle$ over the pair
(or MS path length) distribution function, where $\mu(E)$
is the x-ray absorption coefficient in the absence of disorder.
The effects of disorder and vibrations are additive, but since the factors
due to configurational disorder are dependent on sample history and
preparation, in this paper we focus only on the thermal contribution.
The effect of the DW factors on the XAFS $\chi(k)$ is dominated by the
average over the oscillatory behavior of each path in
the multiple-scattering (MS) path expansion
$\chi_R(k) \propto \sin(2kR+\Phi)$.
If the disorder is not too large the average is conveniently
expressed in terms of the cumulant expansion,\cite{kubo62,crozier88}
\begin{eqnarray}
\label{eq:cum_exp}
\left\langle e^{i2kr} \right\rangle &\equiv& e^{2ikR_0}e^{-W(T)}, \\
   W(T) &=& -\sum_{n=1}^{\infty} \frac{\left(2ik\right)^n}{n!}
  \sigma^{\left(n\right)}(T),
\end{eqnarray}
where $r$ is the instantaneous bond length,
$R_0$ the equilibrium length in the absence of vibrations, and
$\sigma^{\left(n\right)}(T)$ the $n$-th cumulant average. 
For multiple-scattering paths, this length refers to half the total
MS path length.
The dominant effect on XAFS amplitudes comes from the leading exponential
decay factor $W(T)\approx 2k^2\sigma^2(T)$, while
the imaginary terms in $W(T)$ contributes to the XAFS phase. The leading
such contribution is the thermal expansion which comes from the first
cumulant $\sigma^{(1)}(T)$
\begin{equation}
   \sigma^{\left(1\right)} = \left\langle r-R_0 \right\rangle.
   \label{eq:cum_avgs_a} 
\end{equation}
Thus, the mean bond length is
$\langle r \rangle \equiv \bar r = R_0 + \sigma^{\left(1\right)}(T)$.
The skew of the distribution, which is given by the third 
cumulant $\sigma^{(3)}(T)$ contributes a negative phase shift,
and hence the mean distance obtained in fits to XAFS experiment
is typically shorter than that obtained from the first cumulant
alone.\cite{crozier88}
As emphasized above, an accurate account of the effects of anharmonicity
is key to a quantitative treatment of these DW factors over a broad range of
temperatures.
This is illustrated in Fig.\ \ref{fig:cu_wavg}, which shows the strong variation
in the mean phonon frequency $\bar\nu=\bar\omega/2\pi$
\textit{vs} small variations in lattice constant
as calculated using various models described below. 
The expressions for the higher cumulants in Eq. (\ref{eq:cum_exp}) 
simplify when expressed with respect to the mean,
and are \cite{crozier88}
\begin{subequations}
\label{eq:cum_avgs}
  \begin{eqnarray}
    \sigma^{\left(2\right)} &=& \left\langle (r- \bar r)^2 \right\rangle \equiv \sigma^2\left(T\right), \label{eq:cum_avgs_b} \\
    \sigma^{\left(3\right)} &=& \left\langle (r- \bar r)^3 \right\rangle.                                     \label{eq:cum_avgs_c}
  \end{eqnarray}
\end{subequations}

The thermal averages involved in the calculation of the cumulants can be
expressed in terms of the {\it projected vibrational density of states}
(VDOS) $\rho_R(\omega)$.\cite{poiarkova99,poiarkova01,krappe02}
For example, the MSRD $\sigma^2$ for a given 
path $R$ is given by the Debye integral 
\begin{equation}
\label{eq:dw_fac}
  \sigma^2_R(T) = \frac{\hbar}{2\mu_R}
                  \int_{0}^{\infty}{ \frac{1}{\omega}
\coth \Bigl ( \frac{\beta\hbar\omega}{2} \Bigr ) \rho_R\left(\omega\right)}
  ~ d\omega,
\end{equation}
where $\mu_R$ is the reduced mass associated
with the path, $\beta = 1/k_{B}T$, and $\rho_R\left(\omega\right)$ is the
vibrational density of states projected on $R$. In the following,
the path index subscript $R$ is suppressed unless needed for clarity.

The first cumulant
$\sigma^{(1)}$
is generally path-dependent and reflects the anharmonic behavior
of a system.
For monoatomic systems, this quantity is
directly proportional to the net thermal expansion $\Delta a = a(T) -a_0$, which
can be obtained
by minimizing the vibrational free energy $F(a,T)$. 
Within the quasi-harmonic approximation, $F(a,T)$ is given by a  sum
over the internal energy $E(a)$ and the vibrational free energy per atom
\begin{equation}
\label{eq:free_energ}
\begin{split}
   F(a,T) =& E(a) + \\
           &  3 k_B T \int_{0}^{\infty} d\omega\, \ln \left[{
              {2 \sinh \left(\frac{\beta \hbar \omega}{2} \right)} }\right]
\rho_a(\omega) ,
\end{split}
\end{equation}
where $T$ is the temperature, $\rho_a(\omega)$ is the
total VDOS,
and we have assumed cubic symmetry for simplicity.

Furthermore, as pointed out
by Fornasini \textit{et al.},\cite{fornasini04}
the values of the cumulants measured in XAFS experiments 
include two further corrections.
First, perpendicular vibrations lead to a small increase in the
mean expansion observed in XAFS compared to that in x-ray crystallography:
\begin{equation}
\label{eq:s1_corr_perp}
  \Delta \sigma^{\left(1\right)}_{\perp} = \frac{1}{2R_0} \sigma^2_{\perp} .
\end{equation}
We have shown (see Appendix) that $\sigma^2_{\perp}$ and $\sigma^2$
are closely related, and hence that $\sigma^2_{\perp}$ can be
estimated in terms of $\sigma^2$.
Second, the position dependent XAFS amplitude factors 
$\exp (-2R/\lambda ) /R^2$
give rise to an effective radial distribution function
$g(R)\rightarrow g(R)\exp(-2R/\lambda)/R^2$ which shifts 
thermal expansion observed in XAFS by an additional correction
\begin{equation}
\label{eq:s1_corr_lambda}
  \Delta \sigma^{\left(1\right)} = - \frac{2}{R} 
   \left(1 + \frac{R}{\lambda}\right) \sigma^2.
\end{equation}
This second correction is often included in XAFS analysis routines
and has been taken into account in the experimental results presented
here.\cite{fornasini04} Note that 
the corrections in Eq.\ (\ref{eq:s1_corr_perp}) and (\ref{eq:s1_corr_lambda})
are both of the same order of magnitude and partially cancel.

\begin{figure}[t]
\includegraphics[scale=0.35,clip]{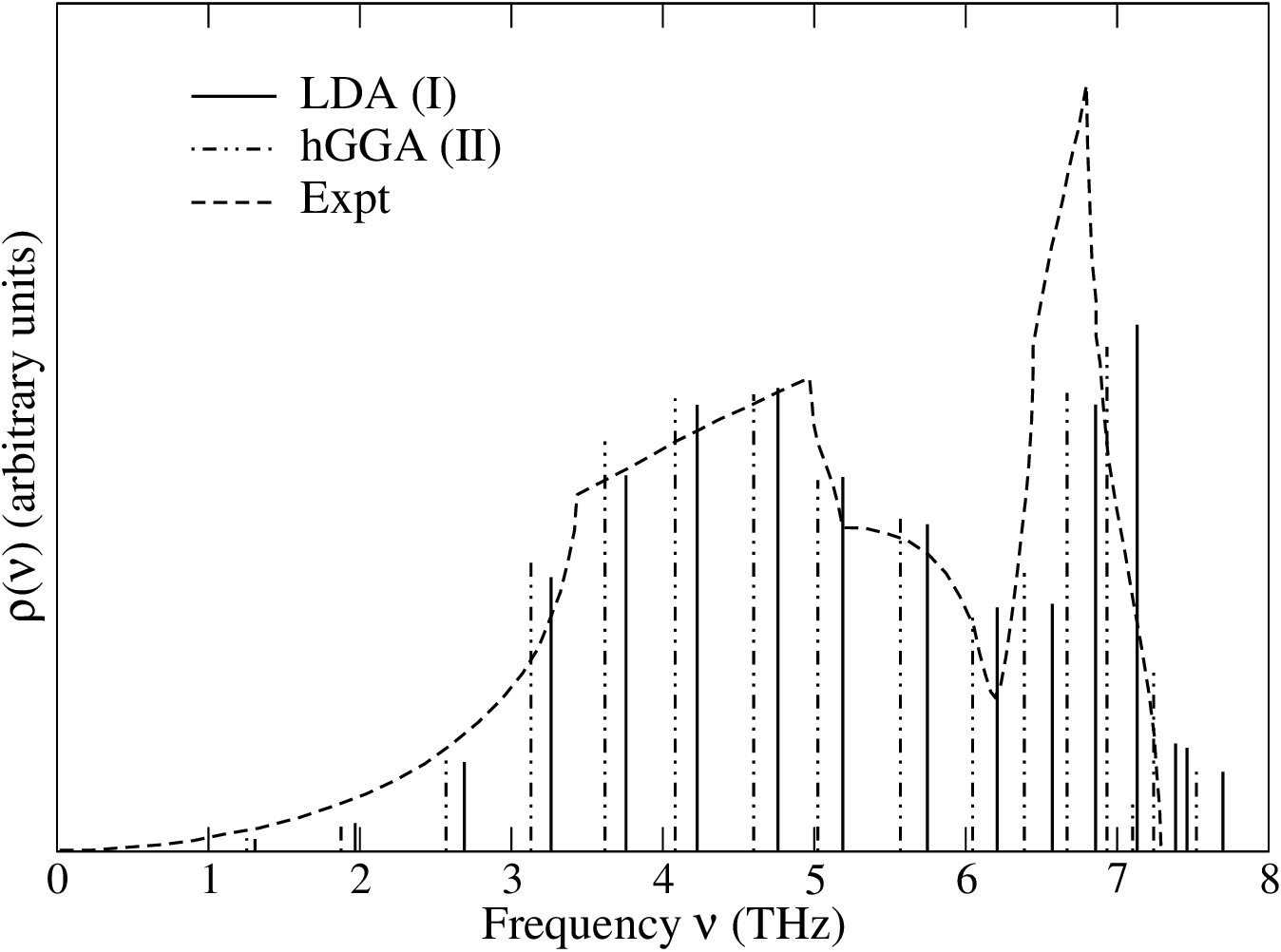}
\caption{\label{fig:cu_vdos}
Total vibrational density of states of Cu at 49 K from DFT
calculations using the LDA with our prescription I; the hGGA
(see text) with our prescription II; and from experiment 
\protect\cite{nicklow67}.
}
\end{figure}

\subsection{Lanczos Algorithm}

The VDOS $ \rho_R\left(\omega\right)$ has
often been approximated by means of Einstein and Debye models based on
empirical data. Although these models are quite useful, especially for
isotropic systems such as metals without highly directional bonds, their
limitations are well known.\cite{dimakis98,poiarkova99}
To overcome some of these limitations
Poiarkova and Rehr \cite{poiarkova99,poiarkova01}
proposed a method in which the VDOS is calculated from the imaginary part
of the lattice dynamical Green's function 
\begin{equation}
\label{eq:greensf}
  \rho_{R}(\omega) = - \frac{2\omega}{\pi}
\mathrm{Im} \Bigl < 0 \Bigl | \frac{1}{\omega^2-{\bf D}+i\epsilon}
\Bigr | 0 \Bigr > .
\end{equation}
Here $\left | 0 \right >$ is a Lanczos seed vector
representing a normalized, mass-weighted initial displacement of the atoms 
along the multiple-scattering path $R$,
and ${\bf D}$ is the {\it dynamical matrix} of force constants
\begin{equation}
\label{eq:rs_dm}
  D_{jl\alpha,j'l'\beta} =
  \left( M_{j} M_{j'} \right)^{-1/2} ~
  \frac{\partial^{2} E(a)}{\partial u_{jl\alpha}\partial u_{j'l'\beta}}.
\end{equation}
where $u_{jl\alpha}$ is the $\alpha = \left\{ 
x,y,z \right\}$ Cartesian displacement of atom $j$ in
unit cell $l$ and $M_{j}$ is 
the mass of atom $j$, and
where $E(a)$ is the internal energy of the system evaluated at
the
lattice constant $a(T)$. Thus, our approach
takes into account the main effects of anharmonicity in terms of
force constants that depend parametrically on the temperature.

Efficient calculations of the lattice dynamical
Green's function can be accomplished using a continued
fraction representation, with parameters obtained 
with the iterative Lanczos algorithm.\cite{deuflhard95} This yields
a many-pole representation for the VDOS which is well suited for
accurate spectral integrations.
The first step in the Lanczos algorithm
corresponds to the {\it correlated Einstein} model,
\begin{equation}
\label{correinsteinmodel}
\rho_R = \delta(\omega-\bar\omega)
\end{equation}
with an Einstein frequency $\bar\omega$ given by
\begin{equation}
\label{baromega}
   \bar\omega^2 = \langle 0 | {\bf D}(R) | 0 \rangle. 
\end{equation}
The frequency $\bar\omega$ corresponds to the rms average over the
projected phonon spectra $\rho(\omega)$.
However, the choice of the Einstein frequency is not unique, and
the appropriate choice depends on the physical quantity being
calculated, as discussed in more detail below.
Poiarkova \textit{et al.} truncated the continued fraction at the second
tier (i.e. second Lanczos iteration), which is usually adequate to converge
the results to about 10\%. Subsequently Krappe and Rossner \cite{krappe02}
showed that at least six Lanczos iterations are required
to achieve convergence to
within $1\%$.  Thus the Lanczos algorithm provides an efficient and accurate
procedure for calculating  MS path-dependent DW factors
from Eq.\ (\ref{eq:dw_fac}). 

The main difficulty in implementing the Lanczos algorithm
lies in obtaining an accurate model 
for the dynamical matrix (or force constants) $\bf D$
for a given system.
Although semi-empirical estimates of interatomic force constants or
Born-von Karman parameters are sometimes available,
their
temperature dependence limits their accuracy and generality.
Similarly, simple models for the vibrational distribution function
(e.g., Einstein and Debye) generally ignore anisotropic behavior.
One of our main aims in this paper is to develop a first principles approach
that allows us to calculate the force constants for various systems
using DFT.  In addition, we have extended the Lanczos algorithm described
above to several other cases
by generalizing the Lanczos seed-state $\left | 0 \right >$. This
allows us to calculate several other quantities including the total
vibrational density of states (VDOS), the vibrational free energy,
thermal expansion, the mean square atomic displacements $u^2(T)$
in crystallographic Debye-Waller factors.\cite{krappe06}
In addition, we calculate $\sigma_{\perp}^2$, 
which yields the perpendicular motion contribution
to the DW factor of Eq.\ (\ref{eq:dw_fac}),
and estimates of the third cumulant.
Representative results for the VDOS calculated by this method
are illustrated in Fig.\ \ref{fig:cu_vdos}.



\begin{figure}[t]
\includegraphics[scale=0.35,clip]{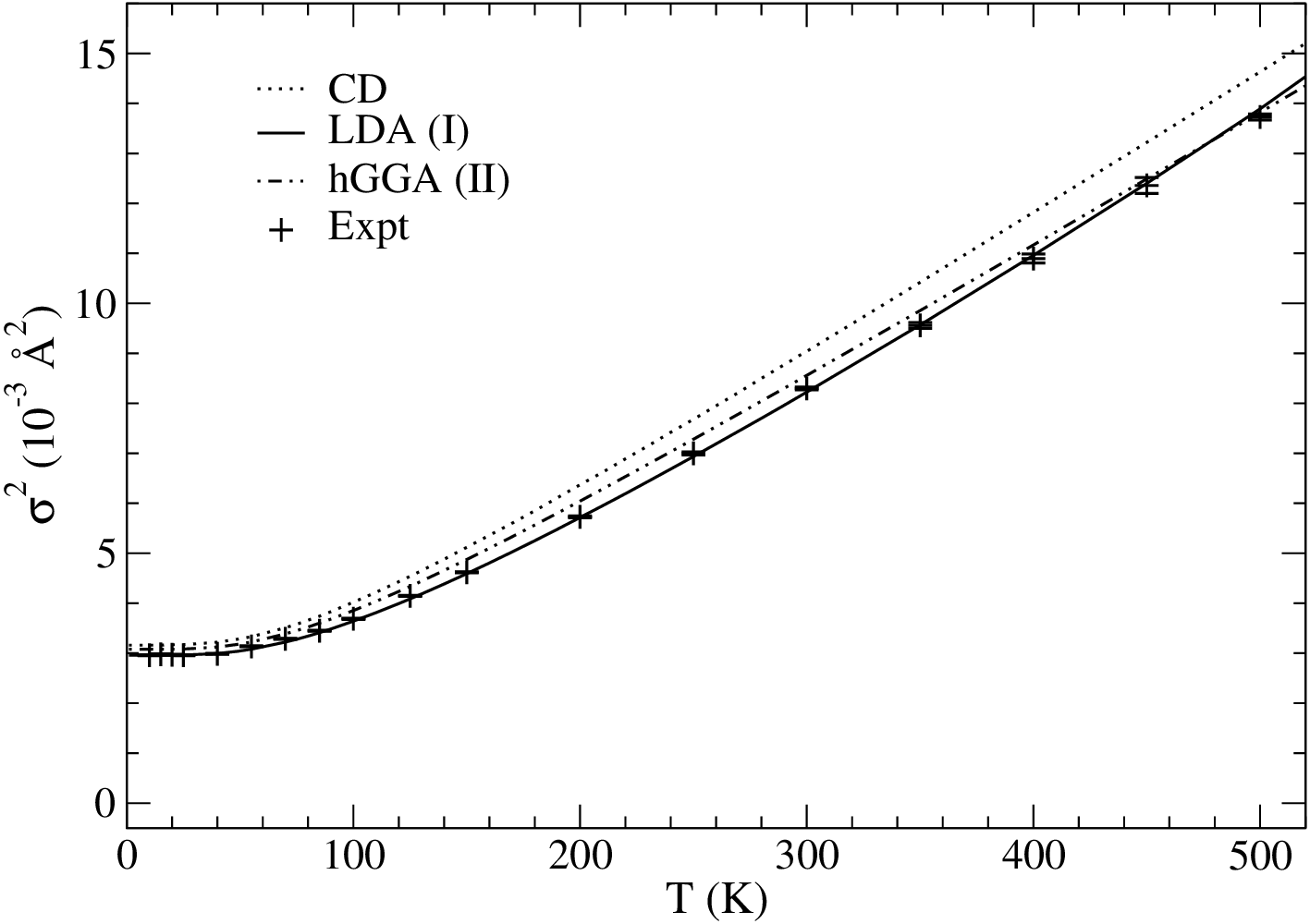}
\caption{\label{fig:cu_dw_t}
         Temperature dependence of the Debye-Waller factor for the nearest
neighbor single scattering path in Cu.  The experimental difference values
\protect\cite{fornasini04} were shifted to match the LDA (I)
results at 0 K.
}
\end{figure}

\subsection{Correlated Einstein Model}
\label{subsec:Corr_Ein_Mod}

Although the cumulants other than the second are often negligible
for small anharmonicity, their calculation using the apparatus of
anharmonic lattice dynamics is computationally demanding.
On the other hand, it has been shown that these cumulants can be approximated 
to reasonable accuracy using a correlated anharmonic Einstein model for each MS
path,\cite{frenkel93,hung97} and this is the method adopted here.
In this approach an Einstein model
is constructed for each MS path keeping only cubic anharmonicity,
yielding the effective one-dimensional potential
\begin{equation}
\label{eq:corr_eins_pot}
  V(x) = \frac{1}{2} k_0 x^2 + k_3 x^3 ,
\end{equation}
where $x$ is the net stretch in a given bond. The Einstein
frequency $\omega_E$ within the quasi-harmonic approximation is then
obtained from the relation Eq.\ (\ref{kk0k3}), i.e.,
$k = k_0+6k_3 \bar x = \mu\omega_E^2$.
This choice of Einstein frequency ensures that the high temperature
behavior of $\sigma^2$ from the Einstein model agrees  with
Eq.\ (\ref{eq:dw_fac}).  The construction of this Einstein model
from the dynamical matrix $\bf D$ along with explicit examples
is given in the Appendix.
The relations between the cumulants for the Einstein model can be used to
obtain estimates for $\sigma^{(1)}$ and $\sigma^{(3)}$. For example, for
the first cumulant 
\begin{equation}
\label{eq:s1_cum_rel}
\sigma^{(1)} = -\eta \frac{3k_3}{k} \sigma^{2}. 
\end{equation}
Note that this relation differs from that in
Refs.\ [\onlinecite{frenkel93,hung97}] in that it contains an extra
multiplicative factor $\eta = 1/\langle \omega^{-2}\rangle \bar\omega^2$,
as discussed in the Appendix.




\section{DFT Calculations }
\label{sec:DFT }

\subsection{Computational Strategy}
\label{sec:meth_dm_aT}

As noted above, one of the main aims 
of this paper is to calculate the force constants
within the quasi-harmonic approximation using DFT
and an appropriate choice of exchange-correlation functional.
Due to the extreme sensitivity of the phonon spectra to the interatomic 
distances, as discussed above,
the most important parameters entering the calculation
of the dynamical matrix
are the lattice constant and the geometry of the system.
A typical example of the effect of
expansion is illustrated in Fig.\ \ref{fig:cu_wavg},
which shows the variation of the first moment of 
the VDOS (i.e. the average frequency $\bar\nu$) projected
along the nearest-neighbor single-scattering  path of Cu.
For comparison Fig.\ \ref{fig:cu_wavg}
also shows $\bar \nu$ obtained  with a model based on
the Born-von Karman  parameters at 298 K.
As expected, when the system expands the vibrational frequencies  are
red-shifted due to the weakening of the interatomic interactions. 
From the common slopes in Fig.\ 1., we see that all of the functionals
have similar Gr{\"{u}}neisen parameters $\gamma\approx$ 2.2 at the
experimental lattice constant 3.61 \AA, in
accord with the experimental value\cite{collins63}
$2.0\pm 0.2$. Note that although at a given lattice constant the GGA functional
always produces a stiffer model than LDA,  i.e., with higher 
mean frequencies, the results at the equilibrium GGA lattice constant
tend to be softer than at the equilibrium LDA lattice constant.  
Moreover, when compared with the experimental value, the LDA and GGA
functionals respectively underestimate and overestimate  the mean frequency
by about 5\%. This translates into a 20-25\% error in the DW factors
calculated with these methods. 
This margin of error is too large to make the DW factors of
significant value in quantitative EXAFS analysis. 
\begin{figure}[t]
\includegraphics[scale=0.35,clip]{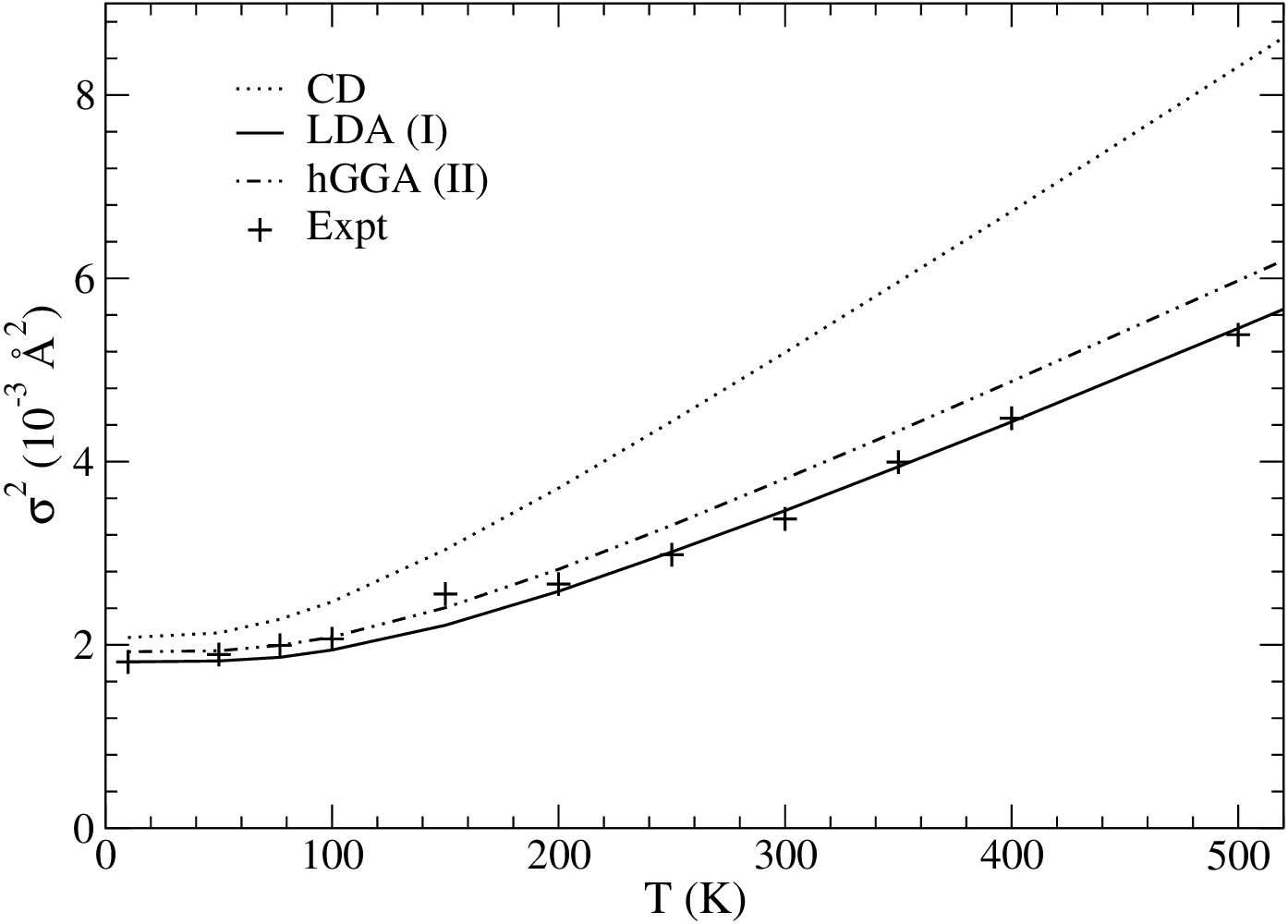}
\caption{\label{fig:ge_dw_t}
         Temperature dependence of the Debye-Waller factor for the nearest neighbor single scattering path in Ge.
	 The experimental difference values \protect\cite{dalba99}
were shifted to match the LDA (I) results at 0 K.
}
\end{figure}
Based on the above considerations, we therefore propose two alternative
prescriptions to stabilize our DW factor calculations:

I.  Our first prescription is based on DFT calculations using the LDA
exchange-correlation functional at the
{\it calculated} equilibrium lattice constants $a(T)$ at a given temperature.
Note however that the errors in the LDA estimates of the lattice constant 
are often larger than those obtained in fits to XAFS experiment. 

II. Our second prescription is based on DFT calculations using a
modified GGA exchange-correlation functional termed hGGA (with half-LDA
and half-GGA) at the {\it experimentally determined}
lattice constant $a(T)$ at a given temperature.
As described below, this functional is constructed on the assumption
that the ``true" functional lies somewhere between pure LDA and GGA.
This second prescription may be useful, for example,
during fits of XAFS data to experiment, during which the interatomic distance
is refined.


Clearly, the use of experimental structural parameters limits prescription
II, since it requires the knowledge of the crystal structure at each of
the temperatures of interest. Such information is only available
{\it a priori} for a handful of systems, although it could be introduced
as part of the fit procedure.

\begin{figure}[t]
\includegraphics[scale=0.35,clip]{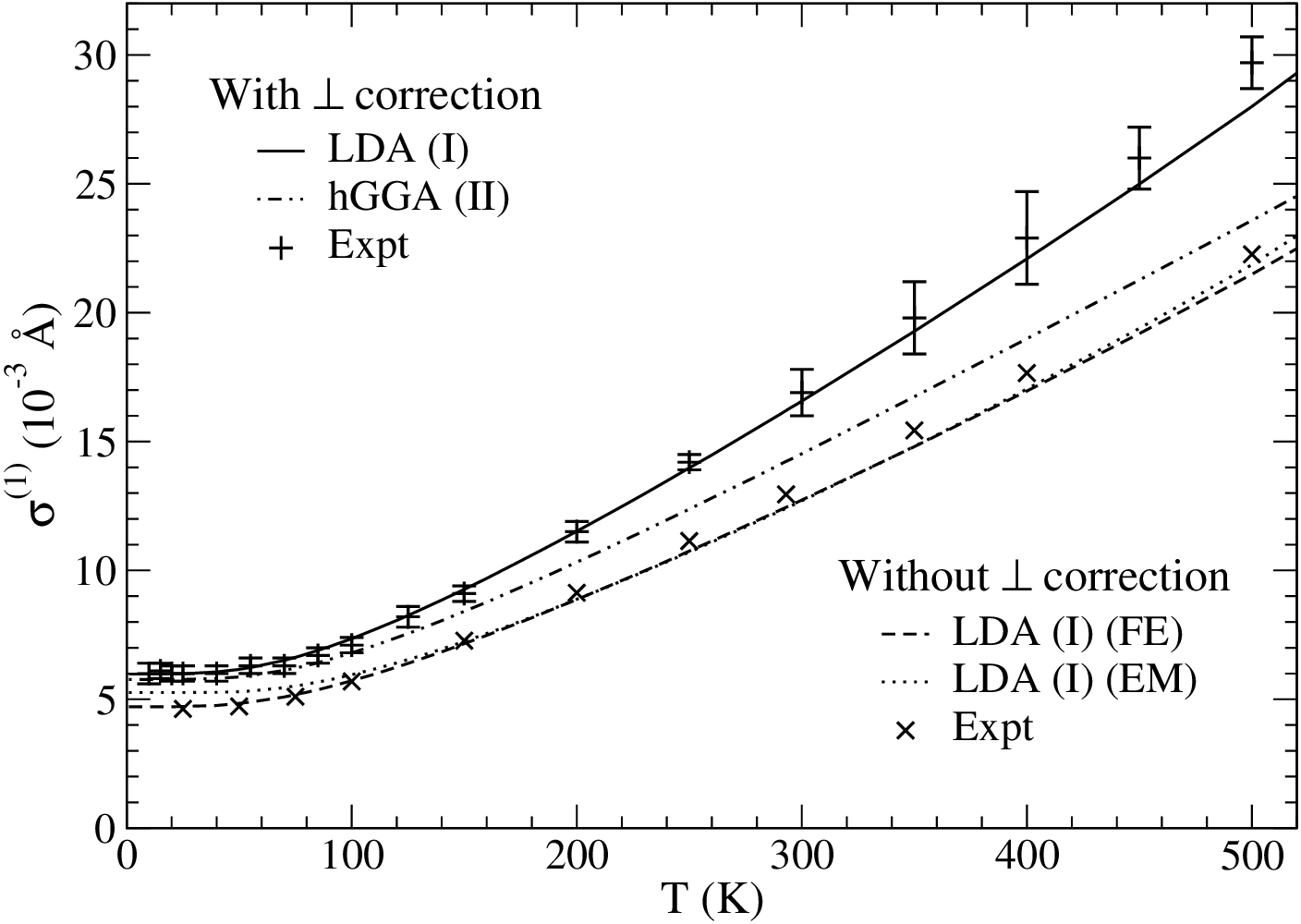}
\caption{\label{fig:cu_c1_t}
         Temperature dependence of the first cumulant for the nearest neighbor single scattering path in Cu, with and without
	 the perpendicular correction from Eq. \ref{eq:s1_corr_perp}, and obtained either from
	 the minimization of the free energy (FE) or from the correlated Einstein model (EM).
	 Both experimental difference values with\protect\cite{fornasini04}
and without\protect\cite{aiphandbook72} the perpendicular motion correction
	 were shifted to match the LDA (I) results at 0 K.
}
\end{figure}

\subsection{Exchange-Correlation Functionals}

In the course of this work, we investigated a number of exchange-correlation
functionals.  Generally, the exchange-correlation functional is attractive 
and hence strongly affects the overall strength and curvature of the
interatomic potential. On the other hand it is well known that LDA
functionals
tend to overbind, yielding lattice
constants smaller than experiment typically by about 1\%. In contrast,
GGA functionals tend to underbind\cite{narasimhan02}
by about the same amount.
These errors are confirmed by our calculations, which show that for Cu
the LDA yields a lattice constant of 3.57 \AA\ at 0 K and 3.58 \AA\ at 298 K,
while the GGA yields 3.69 \AA\ and 3.70 \AA\ respectively, experiment
being 3.61 \AA.
Moreover, the effect of the functionals on the phonon structure 
is even larger.
For example, Narasimhan and de Gironcoli\cite{narasimhan02} show that the
thermal expansion is about 10\% high with LDA and 10\% low with GGA.

\begin{table}
\caption{
Born-von Karman parameters $D^{m}_{ij}$ (N/m)
from neutron scattering compared with
{\it ab initio} calculations from this work.
}
\label{tab:BvK_met}
\begin{ruledtabular}
\begin{tabular}{cccdddf{6}c}
&\tbhc{m}&\tbhc{ij} & \tbhc{LDA}  & \tbhc{GGA}& \tbhc{hGGA} & \tbhc{Expt}&\\
\hline
                &   110  & xx &  14.53 &  11.13 &  13.69 &13.278&[\onlinecite{nicklow67}]\\
\tbh{1}{c}{Cu}  &        & zz &  -3.17 &  -2.18 &  -3.46 &-1.351&\\
\tbh{1}{c}{49 K} &        & xy &  17.12 &  13.12 &  16.52 &14.629&\\
                &   200  & xx &   1.07 &   0.85 &   1.31 &-0.041&\\
                &        & yy &  -0.12 &  -0.11 &  -0.06 &-0.198&\\
\hline
                &   110  & xx & 10.46 &   6.03 &  12.98 &10.71(17)&[\onlinecite{kamitakahara69}]\\
\tbh{1}{c}{Ag}  &        & zz & -3.28 &  -1.65 &  -4.11 & 1.75(20)&\\
\tbh{1}{c}{296 K}&        & xy & 12.48 &   7.27 &  15.94 &12.32(32)&\\
                &   200  & xx &  1.06 &   0.70 &   1.34 & 0.06(29)&\\
                &        & yy & -0.03 &  -0.08 &  -0.05 &-0.23(19)&\\
\hline
                &   110  & xx &  14.22 &   9.17 &  17.97 &16.43(09)&[\onlinecite{kamitakahara69}]\\
\tbh{1}{c}{Au}  &        & zz &  -7.51 &  -5.31 &  -8.78 &-6.54(10)&\\
\tbh{1}{c}{295 K}&        & xy &  18.39 &  12.10 &  23.38 &19.93(14)&\\
                &   200  & xx &   3.96 &   3.30 &   4.42 & 4.04(17)&\\
                &        & yy &  -0.33 &  -0.28 &  -0.25 &-1.27(11)&\\
\hline
                &   110  & xx &  29.17 &  23.67 &  29.94 &25.681(168)&[\onlinecite{dutton72}]\\
\tbh{1}{c}{Pt}  &        & zz &  -7.60 &  -6.67 &  -8.77 &-7.703(251)&\\
\tbh{1}{c}{90 K} &        & xy &  31.44 &  26.02 &  33.50 &30.830(303)&\\  
                &   200  & xx &   4.98 &   4.67 &   5.18 & 5.604(329)&\\
                &        & yy &  -1.56 &  -1.33 &  -1.28 &-1.337(194)&\\
\end{tabular}
\end{ruledtabular}
\end{table}

Although significant effort has been put into so-called meta-GGA
functionals\cite{tao03,staroverov04} that address these issues,
they have not yet been widely implemented.
%
%
Therefore, to be consistent with various results\cite{narasimhan02}
and to preserve the advantages of the LDA and GGA functionals
we have devised a modified functional termed hGGA which is a mixture
of 50\% LDA and 50\% GGA, 
i.e., with a 50\% reduction in both the additional exchange and correlation
terms in the GGA.  The motivation for the 50-50 mixture
stems from the observation that the
experimental values for many quantities are roughly bracketed by the
LDA and GGA predictions.  To simplify the
development, we chose the closely related Perdew-Wang
92\cite{perdew92} (LDA) and Perdew-Burke-Ernzerhof\cite{perdew96} (GGA)
functionals. For this case the equal parts mixing can be achieved with
two simple changes: First, the $\kappa$ parameter in the exchange energy
term in PBE is reduced by half. This change preserves all the conditions
on which PBE was founded, except the Lieb-Oxford bound. Second, the
gradient contribution $H$ to the correlation
energy is also reduced by half.
Similar modified functionals for solids have been proposed by 
Perdew \textit{et al},\cite{perdew07} suggesting that modifications
similar to the hGGA may be more generally applicable.
Fig.\ \ref{fig:cu_wavg} shows the
average frequency obtained with the hGGA functional
and confirms that this yields the desired
behavior. 


\begin{table}
\caption{
    Debye-Waller factors $\sigma^{2}_{n}(T)$ (in $10^{-3}$ \AA$^2$) for
    the single scattering path to the \textit{n}$^{th}$ shell
of some fcc lattice metals.
The experimental difference values were shifted to match the LDA (I) at 80 K and
the experimental error (in parentheses) indicates the error in
the least significant digits.
}
\label{tab:DW_cry_fcc}
\begin{ruledtabular}
\begin{tabular} {ccddddf{5}c}
                    & \tbhc{n} & \tbhc{CD} & \tbhc{LDA(I)} & \tbhc{GGA} & \tbhc{hGGA(II)} & \tbhc{Expt} &   \\
\hline
\tbh{1}{c}{Cu}      & 1 &   6.11  &     5.48  &     6.79  &     5.80  &     5.57(05)  &[\onlinecite{stern80}] \\
\tbhc{190 K}         & 2 &   7.49  &     7.49  &     9.20  &     8.01  &     7.4(3)	 &		 \\
                    & 3 &   7.67  &     7.06  &     8.70  &     7.53  &     6.7(3)	 &			 \\
                    & 4 &   7.76  &     7.02  &     8.68  &     7.48  &     7.0(5)	 &		 \\
\hline
\tbh{1}{c}{Cu} & 1 &   9.04  &     8.22  &    10.45  &     8.56  &     7.99(16)  &[\onlinecite{stern80}] \\
\tbhc{300 K}    & 2 &  11.16  &    11.44  &    14.31  &    12.03  &    11.2(5)	 &		 \\
               & 3 &  11.50  &    10.76  &    13.53  &    11.28  &     9.7(6)	 &		 \\
               & 4 &  11.66  &    10.70  &    13.49  &    11.20  &    11.4(10)   &		 \\
\hline
\tbh{1}{c}{Pt} & 1 &   3.55  &     3.23  &     3.91  &     3.23  &     3.22(05)  &[\onlinecite{stern80}]  \\
\tbhc{190 K}    & 2 &   4.38  &     4.64  &     5.57  &     4.78  &     4.7(3)	 &	       \\
               & 3 &   4.50  &     4.44  &     5.36  &     4.49  &     4.3(4)	 &	       \\
               & 4 &   4.56  &     4.60  &     5.55  &     4.66  &     4.5(4)	 &	       \\
\hline
\tbh{1}{c}{Pt} & 1 &   5.41  &     4.98  &     6.08  &     4.90  &     4.83(05)  &[\onlinecite{stern80}]  \\
\tbhc{300 K}    & 2 &   6.69  &     7.23  &     8.71  &     7.34  &     6.8(5)	 &	       \\
               & 3 &   6.91  &     6.92  &     8.40  &     6.89  &     6.7(6)	 &	       \\
               & 4 &   7.01  &     7.17  &     8.71  &     7.16  &     7.0(6)	 &	       \\
\hline
\tbh{1}{c}{Ag} & 1 &   3.78   &    3.69    &   4.89	&   3.38      &    3.9(3)   &[\onlinecite{newville95}]\\
\tbhc{80 K}     & 2 &   4.56   &    4.97    &   6.57	&   4.60      &    5.4(5)   &		     \\
               & 3 &   4.62   &    4.70    &   5.92	&   4.32      &    4.9(5)   &		     \\
               & 4 &   4.64   &    4.67    &   6.22	&   4.28      &    5.5(5)   &		     \\
\end{tabular}
\end{ruledtabular}
\end{table}




\subsection{Dynamical Matrix}
\label{sec:meth_dm}

The key physical quantity needed in calculations of the Debye-Waller
factors is the dynamical matrix $\bf D$.
With modern electronic structure codes this matrix of force constants
can be calculated with sufficient accuracy from first principles
both for periodic and molecular
systems.\cite{baroni01,lee95,gonze96} In this paper we restrict our
attention to periodic systems
which can be treated, for example, using the methodology
implemented in the ABINIT \cite{gonze02} code,
as described in detail in Ref.\  \onlinecite{gonze97}.
Briefly, the reciprocal space dynamical matrix
\begin{equation}
\label{eq:ks_dm} \tilde{D}_{j\alpha
j'\beta} \left( \mathbf q \right) = \sum_{l'}
 D_{j0\alpha,j'l'\beta} e^{i \mathbf{q} \cdot \left(
\mathbf{R}_{j'l'} -  \mathbf{R}_{j0} \right)}
\end{equation}
is calculated in a $4\times4\times4$ grid of 
${\bf q}$-vectors. This grid is interpolated inside the Brillouin zone and the
real-space  force constants are obtained by means of an inverse Fourier
transform. We find that  such an interpolated grid gives well
converged real-space force constants up to the  fourth or fifth shell. The
neglect of the shells beyond that
introduces an error much smaller than other sources of
error in the method. Finally, since the
calculation of the DW factors uses clusters that typically
include about 20 shells, the full force constant
matrix for these clusters must be built by replicating the $3\times 3$
$D_{jl\alpha,j'l'\beta}$ blocks obtained for each $jl,j'l'$ pair.

\begin{table}
\caption{
    Debye-Waller factors $\sigma^{2}_{n}(T)$ (in $10^{-3}$ \AA$^2$) for the single scattering path for the \textit{n}$^{th}$
shell of some diamond lattice semiconductors.
The experimental difference
values were shifted to match the LDA (I) at 80 K and the
experimental error (in parentheses) indicates the error in
the least significant digits.
}
\label{tab:DW_cry_dia}
\begin{ruledtabular}
\begin{tabular} {ccf{3}f{5}f{3}f{5}f{4}c}
                &  \tbhc{n} &\tbhl{CD}&\tbhl{LDA(I)}&\tbhl{GGA}&\tbhl{hGGA(II)}&\tbhl{Expt}& 		      \\
\hline
\tbhc{Ge}               & 1 &	5.11  &     3.42  &	3.98  &     3.76  &	3.5(1)   &[\onlinecite{stern80}] \\
\tbhc{295 K}             & 2 &	7.43  &    10.38  &    11.91  &     9.70  &	9.6(8)   &		       \\
                        & 3 &	7.64  &    13.09  &    15.03  &    11.84  &              &		       \\
\hline
\tbhc{GaAs}             & 1 &	5.17  &     3.97  &	4.59  &     3.86  &	4.2(1)   &[\onlinecite{stern80}] \\
\tbhc{295 K}             & 2 &	7.75  &    12.70  &    14.28  &    12.01  &    11.7(14)  &		       \\
{\scriptsize (Ga Edge)} & 3 &	7.69  &    14.91  &    16.29  &    14.01  &              &		    \\
\hline
\tbhc{GaAs}             & 1 &	5.15  &     3.96  &	4.59  &     3.86  &	4.2(1)   &[\onlinecite{stern80}] \\
\tbhc{295 K}             & 2 &	7.20  &    10.80  &    11.69  &    10.19  &	9.6(11)  &		       \\
{\scriptsize (As Edge)} & 3 &	7.68  &    14.83  &    16.66  &    14.00  &              &		    \\
\end{tabular}
\end{ruledtabular}
\end{table}

\subsection{Lattice and Force Constants}

The temperature-dependent lattice constant $a(T)$ is obtained by minimizing
$F(a,T)$ in Eq.\ (\ref{eq:free_energ}) with respect to $a$
at a given temperature $T$.  Within  the electronic structure code
ABINIT, the total VDOS $\rho_a(\omega)$ is calculated with histogram 
sampling in $\mathbf{q}$-space. However, we find it more convenient
here to use a Lanczos algorithm in real space,
similar to the approach used for the MSRD. This can be done
by modifying the initial normalized displacement state
$\left | 0 \right >$ in Eq. (\ref{eq:greensf}) to that for 
a single atomic displacement, rather than the displacement along a given
MS path.  If more than one atom is present in the unit cell 
the contributions from each atom must be calculated and added.
Similarly for anisotropic systems one must trace over three
orthogonal initial displacements.  Fig.\ \ref{fig:cu_vdos} shows a typical
VDOS generated using the Lanczos algorithm. We find the free energies
calculated with this approach deviate from the $\mathbf{q}$-space
histogram method by less then 2 meV, i.e., to within 1\%.

To minimize $F(a,T)$ efficiently we proceed as follows: First, the lattice
constant is optimized with respect to the internal energy $E(a)$
and a potential energy surface (PES) for the cell expansion is built around the minimum.
Second, the \textit{ab initio} force constants are computed at each point
of the PES to obtain the vibrational component of $F(a,T)$. Since this
is the most time-consuming part of the calculation, we have taken advantage of
the approximately linear behavior for small variations as illustrated
in Fig. \ref{fig:cu_wavg}. Then, each element of the force constants matrix
is interpolated according to 
\begin{equation}
\label{eq:D_interp}
   D_{jl\alpha,j'l'\beta} =
A_{jl\alpha,j'l'\beta} + B_{jl\alpha,j'l'\beta} ~ \Delta a
\end{equation}
from just two
\textit{ab initio} force constant calculations with slightly different
lattice parameters. 
%
This interpolation scheme allows us to reduce the computational
cost of a typical
calculation by a factor of 2/3, while introducing an error of less
than 2\% in the average frequencies.
Once the values of $F(a,T)$ on the PES are obtained,
we determine the minimum $a(T)$ by fitting $F(a,T)$
to a Morse potential 
\begin{equation}
\label{eq:morse_pot}
F(a,T) = D_0 \left [ e^{-2 \beta (a - a(T))} -2 e^{- \beta (a - a(T))} \right].
\end{equation}
We have estimated that the numerical error
in this minimization is of order $5\times10^{-4}$ \AA\ or less
by fitting only the internal
energy component $E(a)$ and comparing with the minima obtained using
conjugate gradient optimization.

\begin{figure}[t]
\includegraphics[scale=0.35,clip]{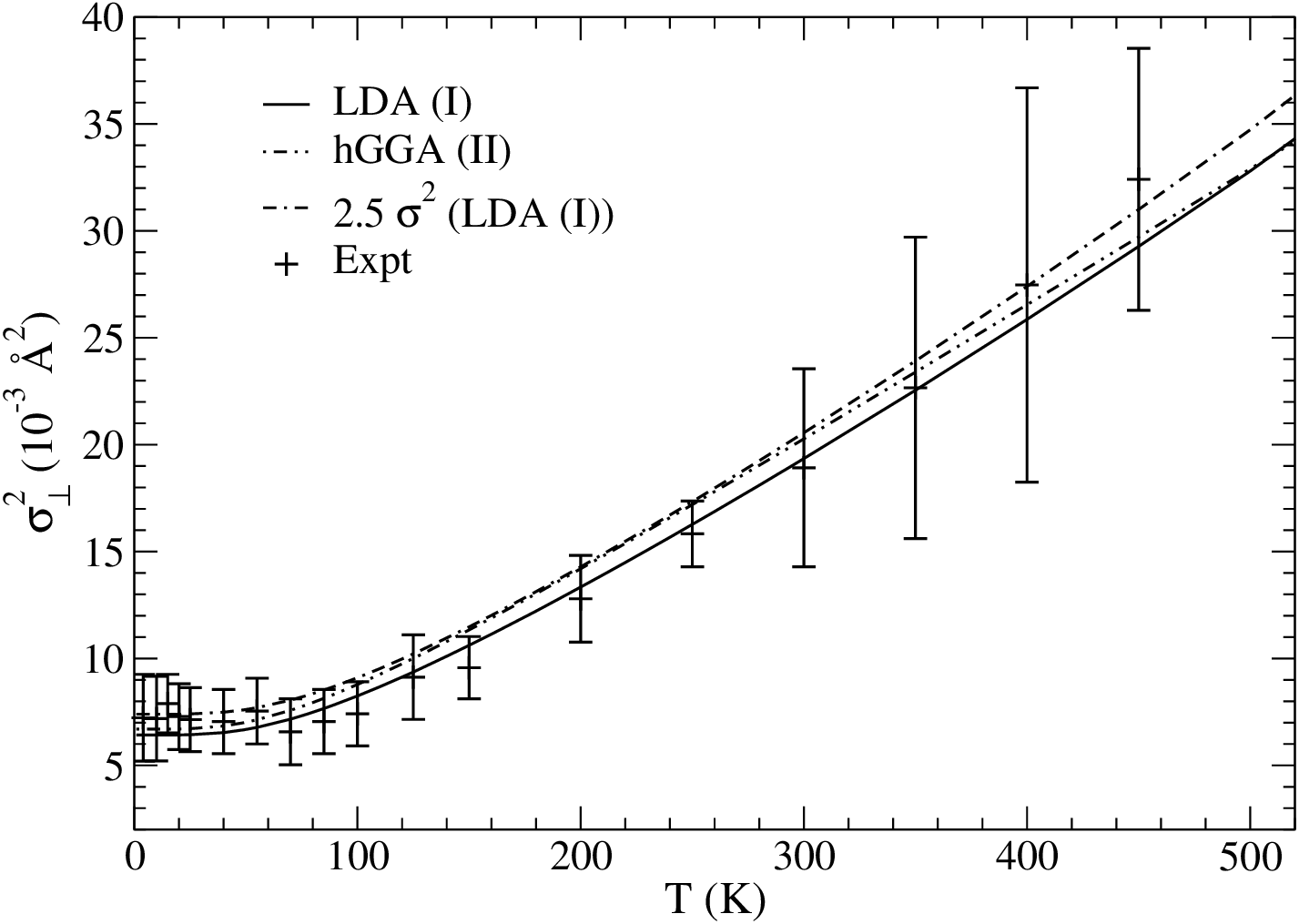}
\caption{\label{fig:cu_s2p_t}
         Temperature dependence of the perpendicular component
$\sigma_{\perp}^2$ of the
Debye-Waller factor for the nearest neighbor single scattering path in Cu.
For comparison we also plot $2.5~\sigma^2$ to show the correlation 
$\sigma_{\perp}^2 \approx \gamma_{\perp} \sigma^2$, together with the
values extracted from experiment.\protect\cite{fornasini04} }
\end{figure}

\subsection{Computational Details}
\label{sec:meth_comp_det}

All the ABINIT calculations reported here use
Troullier-Martins scheme---Fritz-Haber-Institut
pseudopotentials. We found that an $8 \times 8 \times 8$ Monkhorst-Pack
$k$-point grid and an energy
cutoff of 60 au (12 au for Ge) were sufficient to achieve convergence with respect to the DW factors. In
all cases where the geometries were varied, an energy cutoff smearing of 5\% was included to avoid
problems induced by the change in the number of plane wave basis sets. For metallic systems, the
occupation numbers were smeared with the Methfessel and Paxton
\cite{methfessel89} scheme with broadening parameter
0.025. Results are presented for LDA (Perdew-Wang 92\cite{perdew92}) and GGA
(Perdew-Burke-Ernzerhof\cite{perdew96}) functionals, as well as for
our mixed hGGA functional.



\section{Results}
\label{sec:res}

\subsection{Born-von Karman parameters}
\label{subsec:res_BvK}

Phonon dispersion curves are often parametrized in terms of 
so-called  Born-von Karman (BvK) coupling constants. These parameters are
essentially
the Cartesian elements of the real space dynamical matrix defined
in Eq.\ (\ref{eq:rs_dm}). The main difference between the Born-von Karman
parameters and force constants obtained within the quasi-harmonic
approximation is that the former are tabulated at specific temperatures
while the temperature dependence of the quasi-harmonic model arises
implicitly from the dependence of the lattice parameters on
thermal expansion. The  dominant BvK coupling constants
(up to the second neighbor)
are presented in Table \ref{tab:BvK_met}.

We find that both the LDA with prescription I and the hGGA
with prescription II generally give force constants that are within a few
percent of experiment. Typically the LDA force constants with prescription I
are slightly higher than those from the hGGA with prescription II.
Also, note that the transverse components of the BvK
parameters tend to  be overestimated.
We have also considered the pure PBE GGA functionals, but find that they 
produce force constants that are significantly weaker due to their 
larger equilibrium lattice constants (Fig. \ref{fig:cu_wavg}).


\subsection{Mean-square Relative Displacements}
\label{subsec:res_dw}

Calculations of the MSRD
for the dominant first near
neighbor path for fcc Cu are shown in Fig.\ \ref{fig:cu_dw_t}, and
detailed results for various scattering paths
are presented in Table \ref{tab:DW_cry_fcc}.
Both of our prescriptions I and II
yield results in good agreement with experiment. For Cu
even the correlated Debye model is quite accurate.
Note also a slight deviation from linearity in temperature $T$
due to the variation in the dynamical matrix with temperature is 
visible both in the experimental curve and in the calculation
using prescription I.

Similarly, calculations of the MSRD
for the first neighbor path in Ge are shown in Fig.\ \ref{fig:ge_dw_t}, and
detailed results
for various scattering paths are given in Table \ref{tab:DW_cry_dia}.
Again, both of our prescriptions yield
results in good agreement with experiment, with
the LDA prescription being slightly better.
For this case, however, the correlated Debye model
is significantly in error; this is not unexpected given the strong
anisotropy of the diamond lattice. 
Tables \ref{tab:DW_cry_fcc} and \ref{tab:DW_cry_dia}
also include similar results for Ag, Pt and GaAs.


    
%

%


\subsection{Thermal Expansion }
\label{subsec:1st_cum}

The thermal expansion can now be calculated in two ways.
First, by minimizing the free energy of the system using
Eq.\ (\ref{eq:free_energ}) one can
obtain the overall thermal expansion corresponding to the
expansion of the lattice constant $a(T)$. For monoatomic systems the
thermal expansion of any MS path is  simply
proportional to the lattice constant. More generally, the
expansion is MS path dependent, and can be estimated using
the correlated Einstein model of \ref{subsec:Corr_Ein_Mod} and the
Appendix.  From Eq.\ (\ref{eq:s1_cum_rel})
and the Einstein model Gr\"uneisen parameter $\gamma = - k_3R/k$,
this model predicts that the first cumulant $\sigma^{(1)}$
has a temperature dependence proportional to $\sigma^2/R$,
\begin{equation}
\label{eq:s1_cum_rellam}
\sigma^{(1)} = {\frac{3\gamma\eta}{R}} \sigma^{2}, 
\end{equation}
As shown in Fig.\ \ref{fig:cu_c1_t} (dashed and dotted curves),
this correlated Einstein model
estimate for the thermal expansion agrees well with 
that obtained from minimizing the free energy of the system
and with experimental crystallographic data.




\subsection{Perpendicular Motion Contributions}

\begin{figure}[t]
\includegraphics[scale=0.35,clip]{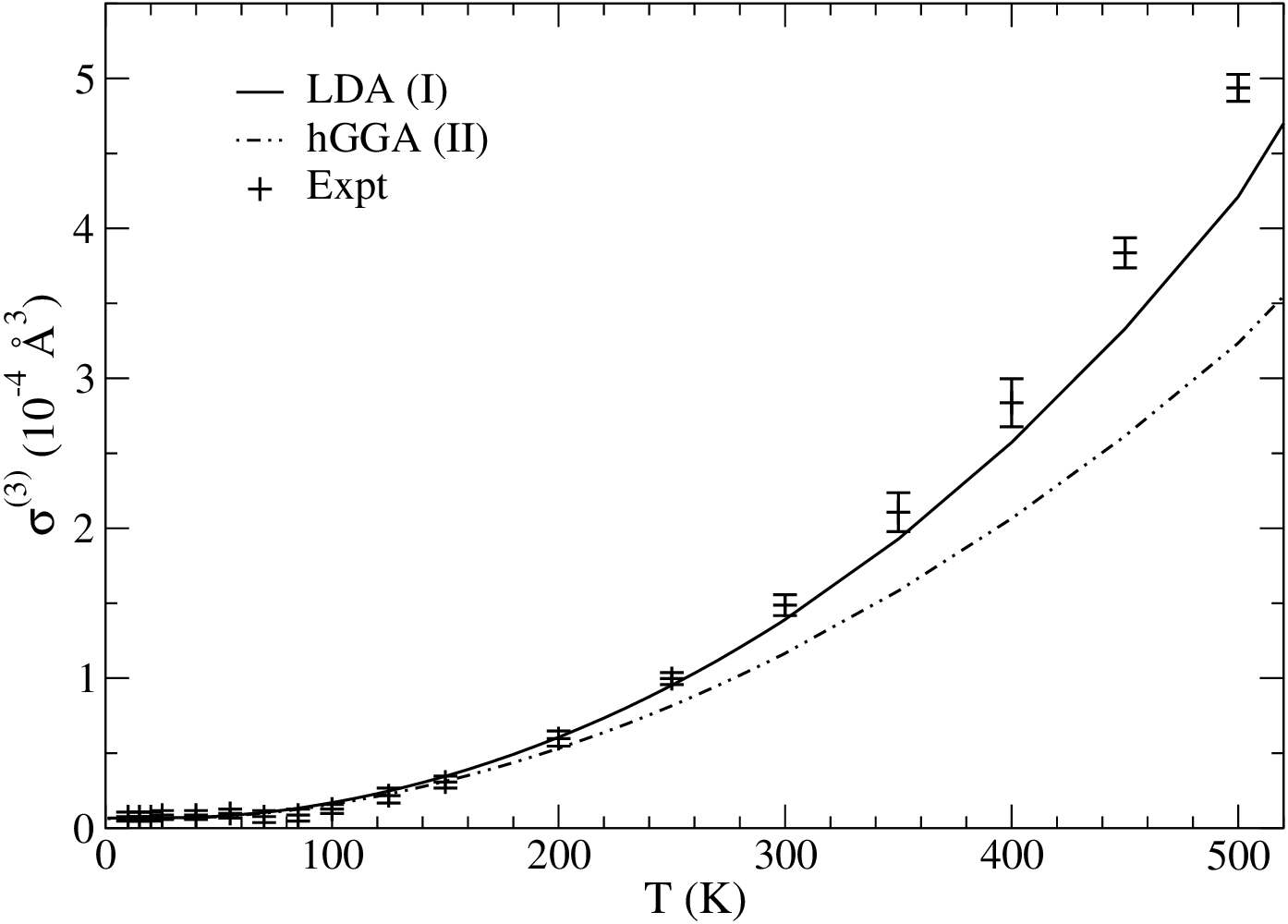}
\caption{\label{fig:cu_c3_t}
         Temperature dependence of the third cumulant for the nearest neighbor
single scattering path in Cu.  The experimental difference
values\protect\cite{fornasini04}
were shifted to match the LDA (I)
results at 0 K.
}
\end{figure}

Fig.\ \ref{fig:cu_c1_t} also shows the first cumulant for Cu obtained by
adding the crystallographic component
$\sigma^{(1)}=\bar x$
and the correction due to perpendicular motion $\Delta \sigma^{(1)}_{\perp}$
from Eq.\ (\ref{eq:s1_corr_perp}).
As observed by Fornasini et al.,\cite{fornasini04} the
mean square perpendicular motion (MSPD) is correlated with $\sigma^2$, i.e.,
$\sigma_{\perp}^2 = \gamma_{\perp} \sigma^2$, with an observed
proportionality constant for Cu
$\gamma_{\perp}\approx 2.5 \pm 0.3$.\cite{fornasini06}
The MSPD can be calculated using our Lanczos procedure with an
appropriately modified seed state $|0\rangle$ for perpendicular vibrations.
This yields a ratio for $\gamma_{\perp}$
that varies from 2.17 to 2.36 between 0 and 500 K, respectively, for Cu.
Moreover, as shown in the Appendix, this ratio can also
be estimated using a correlated Einstein model for fcc structures, and
we derive a value of 2.5 at high temperatures.
The correlated Einstein model also predicts that 
$\gamma_{\perp}$ is weakly temperature-dependent,
reducing to about $\sqrt 5\approx 2.24$ at low temperature.
We also show that for fcc structures the correction due to perpendicular motion
is smaller than the crystallographic contribution by a factor of
$\gamma_{\perp}/6\gamma$, which for Cu is about 20\%.
%
%
To illustrate this correlation, Fig. \ref{fig:cu_s2p_t} shows the 
perpendicular motion contribution $\sigma_{\perp}^2$ calculated
both by the Lanczos procedure and with a constant correlation
factor $\gamma_{\perp}=2.5$.

%
We have carried out similar calculations of $\sigma_{\perp}^2$
for the case of diamond lattices. Due to the strongly directional
bonding in diamond structures, and non-negligible bond bending
forces, the calculations are more complicated than for fcc materials.
Our {\it ab initio} calculations
using the LDA with prescription I yield a ratio that varies from
3.4 to 7.2 between 0 and 600 K, in reasonable agreement with experiment
where $\gamma_{\perp}$ varies between 3.5$\pm$0.6 and 6.5$\pm$0.5 in the
same range.  \cite{fornasini06}
In contrast our single near neighbor spring model (Appendix) gives
a smaller high temperature value $\gamma_{\perp}=3.5$ and the
addition of a single bond-bending parameter does not improve the
agreement.

\subsection{Third Cumulant }
\label{subsec:3rd_cum}

As for the first cumulant, the third cumulant can be estimated from
the correlated Einstein model, and the relation
\begin{equation}
\label{eq:s3_cum_rel}
\sigma^{(3)} =\eta \left[ 2 - \frac{4}{3}
\left(  \frac{\sigma^{2}_{0}}{\sigma^{2}} \right)^2 \right]
             \sigma^{(1)} \sigma^{2}.
\end{equation}
Again an additional scaling factor $\eta$ 
is needed to correct the original Einstein model relations
when $\sigma^{(1)}$ and $\sigma^2$ are replaced by the full results
from our LDA calculations.  Also the presence of this factor $\eta$
gives another correction to the relation
$\sigma^{(1)} \sigma^{2}/\sigma^{(3)}\approx 2$ given by
classical models\cite{stern91}
or the correlated Einstein model at high temperatures.

%

\begin{figure}[t]
\includegraphics[scale=0.35,clip]{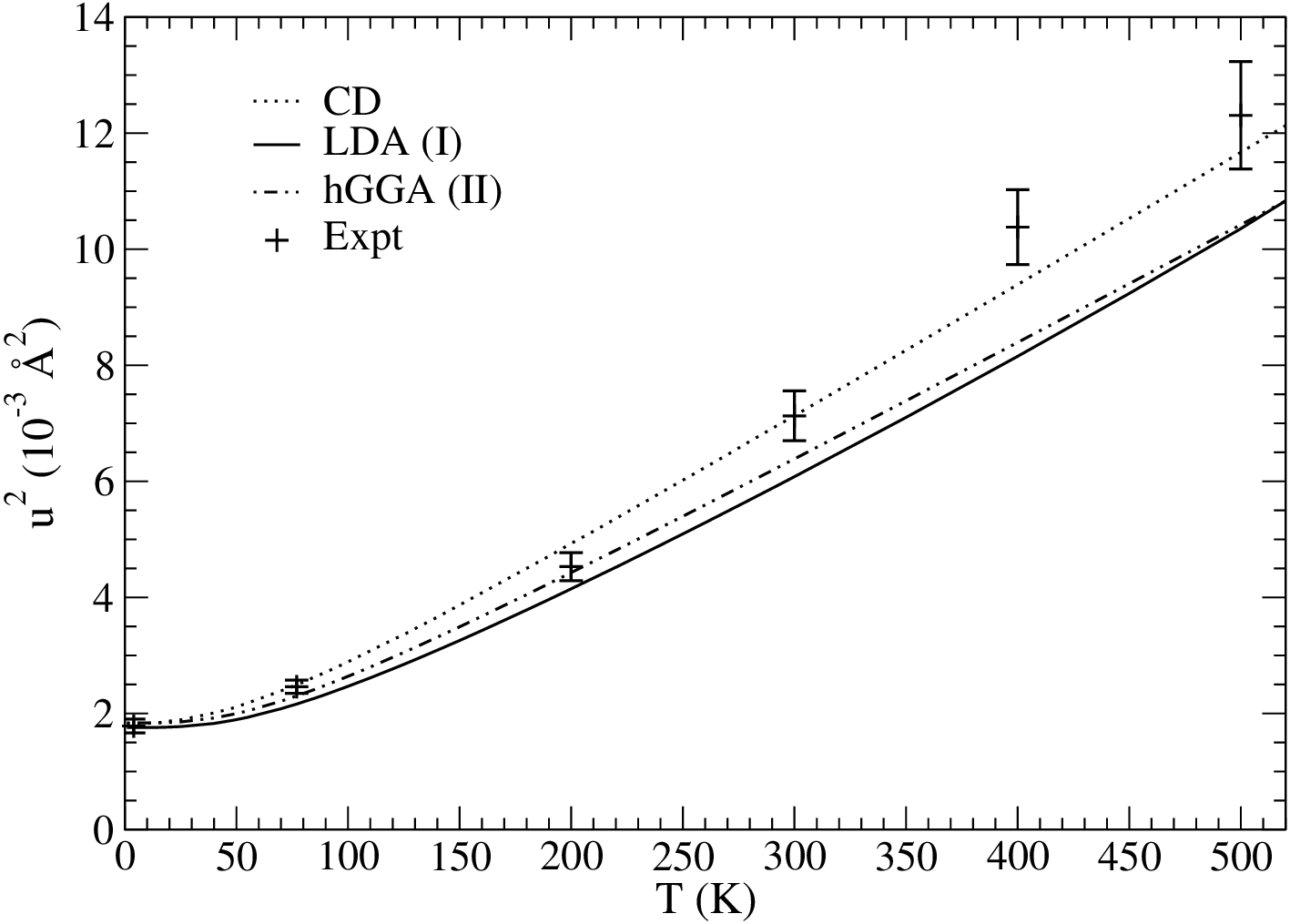}
\caption{\label{fig:cu_u2_t}
	 Temperature dependence of the crystallographic Debye-Waller factor
for the nearest neighbor single scattering path in Cu, and compared
to experimental values \protect\cite{flinn61}.
}
\end{figure}

\subsection{Crystallographic Debye-Waller Factors}
\label{subsec:crystalDW}

Finally, we present results for the x-ray and neutron crystallographic
Debye Waller factors $W(T)=(1/2) k^2 u^2(T)$, where 
the mean-square displacements
$u^2(T)=\langle ({\bf u\cdot\hat{\bf R}}{})^2\rangle$
are given by Eq.\ (\ref{eq:dw_fac}), with  $\rho_a(\omega)$
given by the total vibrational density of states per site,
as calculated by our Lanczos algorithm with an appropriate
seed state.\cite{krappe06}
For this case good agreement is obtained for both of our DFT prescriptions
at low temperature, although the errors become more significant
at higher temperatures. Also, we find that the convergence of the
Lanczos algorithm is slower than for the path dependent Debye-Waller
factors, requiring approximately 16 iterations to achieve convergence to 1\%.

\section{Discussion and Conclusions}
\label{sec:concl}

We have developed a first principles approach for calculations of the
Debye-Waller factors in various x-ray spectroscopies, based on
DFT calculations of the dynamical matrix and phonon spectra for a given
system. We find that the results depend strongly on the choice
of exchange-correlation potential in the DFT, but we have developed two
prescriptions that yield stable results, 
one based on the LDA and one based on a modified GGA termed hGGA.
Calculations for the crystalline systems presented here show that our 
LDA prescription yields good agreement with experiment
for all quantities, typically within about $\pm$ 10\%.
Second, if the lattice constant is known
{\it a priori}, our hGGA prescription also provides an accurate
procedure to estimate the MSRD.
Anharmonic corrections and estimates of the contribution from perpendicular
vibrations are estimated using a correlated Einstein model.
For these anharmonic quantities, however, we have found that the comparative
softness of the lattice dynamics with the GGA and hGGA functionals leads to
results which are somewhat less accurate than those for the LDA.
Finally we have also calculated the crystallographic Debye Waller factors.
Our approach also yields good results
for calculations of DW factors in anisotropic systems, as illustrated for
for Ge and GaAs.  All of these results demonstrate that the prescriptions
developed herein can yield quantitative estimates of Debye-Waller factors
including anharmonic effects in various crystalline systems, and generally
improve on phenomenological models.  Extensions to molecular systems are
in progress.

%
%
%


\begin{acknowledgments}
We thank S. Baroni, K. Burke, A. Frenkel, X. Gonze,
N. Van Hung, P. Fornasini, M. Newville, and J. Perdew for many 
comments and suggestions. This work is supported in part by the DOE Grant
DE-FG03-97ER45623 (JJR) and DE-FG02-04ER1599 (FDV), and was facilitated by
the DOE Computational Materials Science Network.
\end{acknowledgments}

\appendix*
\section{}

In this Appendix we briefly discuss the correlated Einstein model
used in estimating anharmonic contributions to the DW factors.
The model is illustrated with an application to the correlated Einstein
model for calculating the mean-square radial displacement (MSRD)
$\sigma^2$ and mean square perpendicular displacement (MSPD)
$\sigma_{\perp}^2 = \langle |\Delta \vec u_{\perp}|^2 \rangle$.

The construction of Einstein models is not unique in that different
physical quantities reflect different averages over the VDOS.
For example, the theoretical MSRD given by Eq.\ (\ref{eq:dw_fac})
reflects an average over a thermal weight factor
varying as $1/\omega^2$ at high temperatures. 
Thus the Einstein model parameters in our prescription are
constructed to preserve the correct high temperature behavior
of $\sigma^2$.\cite{sevillano79,crozier88}
The first step in this construction is the
calculation of $\bar\omega^2$ from the
total potential energy for a net displacement
$x$ of a path along a particular seed displacement state $|0\rangle$.
Next this value is renormalized to give the correct MSRD at
high temperatures.  Thus we define
\begin{equation}
  k_0= \eta\mu \langle 0|{\bf D}|0\rangle
\end{equation}
where $\bar\omega^2$ is given from  Eq.\ (\ref{baromega}) and
the factor $\eta = 1/\langle \omega^{-2}\rangle \bar\omega^2$,
where $\langle \omega^{-2}\rangle$ is the inverse second moment
of the projected VDOS.
The cubic coupling $k_3$ is then defined to be consistent with the
variation in $k$ given by the Gr\"uneisen parameter 
\begin{equation}
\gamma = \frac{d \ln \bar\omega}{d \ln {R^3}} = -\frac{k_3R}{k}.
\end{equation}
and hence must be is similarly renormalized
\begin{equation}
\label{eq:k3}
   k_3 = \frac{\eta}{6} \frac{d}{dR} \langle 0 |  {\bf D}(R) | 0 \rangle .
\end{equation}
Then the Einstein frequency $\omega_E$ in the quasi-harmonic approximation is
obtained from the relation\cite{frenkel93}
\begin{equation}
\label{kk0k3}
  k = k_0+6k_3 \bar x = \mu\omega_E^2.
\end{equation}
where $\mu$ is the reduced mass.
For Cu using the LDA (I) prescription for the dynamical matrix,
this procedure yields $\eta = 0.73$, $k_0=54.7$ N/m,
$k_3=-48.4$ N/m\AA, and $k= 51.1$ N/m.

The scaling factor $\eta$ thus forces the relation
$\sigma^2(T) \rightarrow \sigma_E^2(T) = k_BT/\mu\omega_E^2$ at
high temperature, where $\sigma^2(T)$ in the Einstein model is
\begin{equation}
\label{em_sigma2}
\sigma_E^2(T) = \sigma^{2}_{0} \coth
\left(\frac{\beta\hbar\omega_E}{2}\right),
\end{equation}
and the zero-point value $ \sigma_0^2= \sigma^2(0) = \hbar\omega_E/2 k$.

Then, from $\sigma^2$ and relations between the
cumulants,\cite{frenkel93,hung97} one can obtain
MS path-dependent estimates for $\sigma^{(1)}$ and $\sigma^{(3)}$.
When these relations are expressed in terms of the calculated
(or experimental) values of the cumulants, we have found it necessary
to include multiplicative factors of $\eta$ compared to the pure Einstein model
expressions,\cite {frenkel93,hung97}   to obtain quantitative agreement with
experimental results e.g., as shown in Figs.\ \ref{fig:cu_c1_t}
and \ref{fig:cu_c3_t}.


As a second example, we construct such a model for monoatomic
fcc Cu starting from
an anharmonic pair potential. That is, we will assume that the lattice
dynamics can be described by an anharmonic pair potential $V_0$
between near-neighbor bonds of the form
\begin{equation}
\label{eq1}
 V_0(x)  \cong  \frac{1}{2}k^0x^2 + k^0_3x^3 .
\end{equation}
Here $x$ is the net displacement $x$ along the
bond direction, with positive displacements referring to expansion
and negative to compression. \cite{frenkel93,hung97}

 First consider the potential energy $V_{\parallel}(x)$
for vibrational displacement $x$ along
the bond $(0R)$ between lattice points $(0,0,0)$
and $\vec R=R(0,1,1)/\sqrt{2}$.
The net anharmonic potential $V_{\parallel}(x)$ is then given by
Eq.\ (\ref{eq:corr_eins_pot})
with a displacement state $x|0\rangle$ defined by $\vec u_0 = (x/2)
(0,1,1)/\sqrt{2}$, and $\vec u_R = (-x/2) (0,1,1)/\sqrt{2}$.
Then constructing the dynamical matrix using Eq.\ (\ref{eq1}) with
small displacements, we find
a net spring constant $k_0 = \eta\langle 0| D | 0\rangle = (5\eta/2) k^0$,
in agreement with Ref.\ \onlinecite{hung97}.
This result can alternatively be obtained
by summing the 23 pair potentials between the
shared bond $(0R)$, the 11 nearest neighbor bonds to the origin
and the 11 other nearest neighbor bonds to $R$, giving
$V_{\parallel}(x) = V_0(x) + 2 V_0(-x/2) + 8 V_0(x/4) + 8 V_0(-x/4) + 4
V_0(0)$. Similarly we find that the anharmonic coupling
[cf.\ Eq.\ (\ref{eq:k3})] is
 $k_3 = (3\eta/4) k_3^0$,
so that
\begin{equation}
\label{eq4}
 V_{\parallel}(x) = \frac{1}{2}\left(\frac{5\eta}{2}  k^0\right)
x^2 + \left(\frac{3\eta}{4} k^0_3\right) x^3,
\end{equation}
where we have again included a factor $\eta$ so that the Einstein
model for $\sigma^2$ agrees with the expression from the inverse
second moment of the VDOS.

In a similar way, we can develop a correlated Einstein model to describe the
perpendicular vibrations. For this case we consider a vibrational displacement 
$\Delta \vec u_{\perp}$ of length 
$y = |\Delta \vec u_{\perp}|$ {\it perpendicular} to the bond between
$(0,R)$. Thus we set $\vec u_0 = (y/2) (0,1,-1)/\sqrt{2}$
and $\vec u_R = (-y/2) (0,1,-1)/\sqrt{2}$, where $R$ is the
nearest neighbor distance. The net potential $V_{\perp}(y)$ is again
obtained by summing the 23 pair potentials between the
shared bond $(0R)$, the 11 nearest neighbor bonds to the origin
and 11 others to $R$, similar to the calculation above for vibrations
along the bond. For this case, two bonds are stretched by $y/2$,
two contracted by $y/2$, three unchanged, eight stretched by $y/4$,
and eight contracted by $y/4$, yielding a net sum
$ V_{\perp}(y) = 2V_0(y/2) + 2V_0(-y/2)+8V_0(y/4)+8V_0(-y/4)+3V_0(0)$,
and hence
\begin{equation}
\label{eq3}
 V_{\perp}(y) = \frac{1}{2} (2\eta k^0) y^2.
\end{equation}
Note that by symmetry, the net cubic anharmonic contribution vanishes.
Thus the effective spring constant for the MSPD is $k=2\eta k^0$
and predicted to be insensitive to thermal expansion.
The correlated Einstein model $V_{\perp}(z)$
is clearly the same for the MSPD along the $z$-axis,

With these results we can show that the MSPD for the first neighbor path
in fcc materials $\sigma_{\perp}^2$ is correlated with
$\sigma^2 = \langle |\Delta \vec u_{\parallel}|^2 \rangle$.
Both the MSPD and MSRD in the Einstein model are given by 
Eq.\ \ref{em_sigma2},
with their respective Einstein frequencies $\omega_E=(k/\mu)^{1/2}$.
For the total contribution from perpendicular vibrations, one has to
multiply by two
to get the net $\sigma_{\perp}^2$ from both independent axes.
At high temperatures for example, we obtain the MSPD
$\sigma_{\perp}^2=2 k_B T/2\eta k^0$.  This is higher than the MSRD
$\sigma^2 = (2/5) k_BT/\eta k^0$ by a factor of $\gamma_{\perp}=5/2.$
The model also predicts
a weakly temperature dependent ratio
\begin{equation}
\label{eq6}
\gamma_{\perp} (T) = {\frac{\sigma_{\perp}^2(T)}{\sigma^2(T)}} = 2
\frac{\omega_E\coth (\beta\hbar\omega_E^{\perp}/2)}
     {\omega_E^{\perp} \coth (\beta \hbar\omega_E/2) }.
\end{equation}
This ratio varies between $\sqrt5 = 2.236$ and 2.5
with increasing temperature.
 Thus the ratio $\gamma_{\perp}$ obtained with the correlated Einstein model,
for the fcc lattice depends only on geometry and describes the anisotropy
of the vibrational ellipsoid
in monoatomic fcc structures reasonably well.

Because of the above relation between $\sigma_{\perp}^2$
and $\sigma^2$ in the Einstein model, the perpendicular motion correction
can be related to the contribution to lattice expansion from anharmonicity.
Thus from Eq.\ (\ref{eq:s1_corr_perp}) and
Eq.\ (\ref{eq:s1_cum_rellam}), we find that
\begin{equation}
{\Delta \sigma^{(1)}_{\perp} }/{\sigma^{(1)}} = {\gamma_{\perp}}/{6 \eta\gamma}.
\end{equation}
%
For fcc Cu this ratio predicts a correction to the first cumulant
$\sigma^{(1)}$ from perpendicular motion
of about 25\%. Indeed, this shift is comparable to the observed differences
in the thermal expansion with and without the perpendicular motion
correction observed in Fig.\ {\ref{fig:cu_c1_t}.
Thus for the dominant near neighbor bonds, the correlated Einstein model
predicts a comparatively small but non-negligible effect of perpendicular
motion on EXAFS distance determinations.


\end{document}